\documentclass[pra,nofootinbib, twocolumn,floatfix,superscriptaddress]{revtex4-2}  
\expandafter\let\csname equation*\endcsname\relax
\expandafter\let\csname endequation*\endcsname\relax

\usepackage{graphicx}  
\usepackage{orcidlink} 
\usepackage{amsfonts, amssymb}   
\usepackage{bm}        
\usepackage{amsmath}    
\usepackage{amsthm}    
\usepackage{verbatim}   
\usepackage{xcolor}      
\usepackage{hyperref}   
\usepackage[caption=false]{subfig}

\usepackage{float}
\usepackage{physics}
\usepackage{enumerate}
\usepackage{mathrsfs}
\usepackage{relsize}
\usepackage{mathtools}
\usepackage{wasysym}
\usepackage{tikz}
\usetikzlibrary{graphs, arrows,chains,matrix,positioning,scopes,shadows}
\usepackage{scalerel,stackengine}
\usepackage{accents}
\usepackage{natbib}

\stackMath
\newcommand{\stretchedhat}[1]{%
\savestack{\tmpbox}{\stretchto{%
  \scaleto{%
    \scalerel*[\widthof{\ensuremath{#1}}]{\kern.1pt\mathchar"0362\kern.1pt}%
    {\rule{0ex}{\textheight}}
  }{\textheight}%
}{2.4ex}}%
\stackon[-6.9pt]{#1}{\tmpbox}%
}
\newcommand{\stretchedtilde}[1]{%
\savestack{\tmpbox}{\stretchto{%
  \scaleto{%
    \scalerel*[\widthof{\ensuremath{#1}}]{\kern.1pt\mathchar"307E\kern.1pt}%
    {\rule{0ex}{\textheight}}
  }{\textheight}%
}{2.4ex}}%
\stackon[-6.9pt]{#1}{\tmpbox}%
}
\newtheorem{theorem}{Theorem}

\newtheorem{lemma}{Lemma}

\theoremstyle{definition}

\definecolor{rkrPurple}{HTML}{73024F}

\begin{document}


\title{Note on Von Neumann Entropy and the Ordering of Inverse Temperatures} %

\author{Rohit Kishan Ray\orcidlink{0000-0002-5443-4782}}
\email{rkray@ibs.re.kr}
\affiliation{Center for Theoretical Physics of Complex Systems, Institute for Basic Science (PCS-IBS), Daejeon - 34126, South Korea}

\begin{abstract}
I show that for two inverse temperatures \(\beta_1\) and \(\beta_2\), the von Neumann entropy \(S(\rho_\beta)\) of the Gibbs state \(\rho_\beta\) for a given Hamiltonian \(H\) satisfies $S(\rho_{\beta_1}) \geq S(\rho_{\beta_2}) \iff \beta_{1} \leq \beta_{2}$. That is, von Neumann entropy is a monotonically increasing function of temperature.
\end{abstract}

\maketitle
The von Neumann entropy is an important aspect of quantum theory \cite{vonneumann_2018_mathematical}. For a thermal state or Gibbs state \(\rho_\beta\) associated with a Hamiltonian \(H\) with inverse temperature \(\beta\), the von Neumann entropy is given by:
\begin{equation}
    S(\rho_\beta) = -\Tr\left(\rho_\beta \log\rho_\beta\right),
\end{equation}
where the density matrix is given by:
\begin{equation}\label{eq:density}
    \rho_\beta = \dfrac{e^{-\beta H}}{\Tr\left(e^{-\beta H}\right)}.
\end{equation}
We consider, the Hamiltonian \(H\) has eigenvalues \(E_i\) ordered in a non-decreasing fashion.
Although, it is intuitive from classical thermodynamics that for a given system entropy increases with temperature, a rigorous proof of the same in case of discrete system scarce in literature. Herein, I am giving the following proof to the relation (Eq. (\ref{eq:entropy_maj}) below. We claim --- 
\begin{align}\label{eq:entropy_maj}
    S(\rho_{\beta_1}) \geq S(\rho_{\beta_2}) \iff \beta_{1} \leq \beta_{2}\, ,
\end{align}
This simple proof is based on majorization theory \cite{marshall_2011_inequalities}. We demonstrate that the eigenvalue distribution of a lower-temperature thermal state majorizes that of a higher-temperature state. Given that the von Neumann entropy is Schur concave \cite{kvalseth_2022_entropy}, this immediately implies the desired monotonicity result. This proof is purely algebraic and holds for all finite-dimensional quantum systems. 

Let us consider the following lemma. 
\begin{lemma}\label{lemma:ratio}
    If the for all positive reals $p_i,q_i$, $i=1(1)N$ we have the following relation:
    \begin{equation}\label{eq:lemma_1}
        \dfrac{p_1}{q_1}\geq \dfrac{p_i}{q_i} \geq \dfrac{p_N}{q_N},
    \end{equation}
    then it implies,
    \begin{equation}\label{eq:lemma_2}
        \dfrac{\sum_{i=1}^k p_i}{\sum_{i=1}^k q_i} \geq \dfrac{\sum_{i=1}^N p_i}{\sum_{i=1}^N q_i}, ~~~~~~ \text{for }~k=2(1)N-1
    \end{equation}
\end{lemma}
\noindent\textit{Proof:} Using the identity, if $\frac{p}{q}\geq \frac{r}{s}$ for $p,q,r,s \geq 0$ and real (the property of arithmetic mean),
\begin{equation}\label{eq:proof_lemma_identity}
    \frac{p}{q}\geq \frac{p+r}{q+s} \geq \frac{r}{s}\,,
\end{equation}
we proceed to use proof by induction. \\
\noindent\textit{Step 1:} $k=2$, we have using the above relation in Eq. (\ref{eq:proof_lemma_identity})
\begin{equation}\label{eq:step 1}
    \dfrac{p_1}{q_1}\geq \dfrac{p_1+p_2}{q_1+q_2} \geq \dfrac{p_2}{q_2}.
\end{equation}
\noindent\textit{Step 2:} We take $k=3$, and using the statement of the Lemma \ref{lemma:ratio}), we have
\begin{align}
    \dfrac{p_2}{q_2} & \geq \dfrac{p_3}{q_3},~~~\text{from Eq. (\ref{eq:lemma_1}} \notag \\
    \implies        \dfrac{p_1+p_2}{q_1+q_2} & \geq \dfrac{p_3}{q_3}, ~~~\text{from Eq. (\ref{eq:step 1})}\notag                                       \\
    \implies   \dfrac{p_1+p_2}{q_1+q_2} & \geq  \dfrac{p_1+p_2+p_3}{q_1+q_2+q_3} \geq\dfrac{p_3}{q_3}. 
\end{align}
\noindent\textit{Step 3:} Lets assume for some $k=r<N-1$ the following holds true,
\begin{equation}
    \dfrac{\sum_{i=1}^{r-1} p_i}{\sum_{i=1}^{r-1} q_i} \geq \dfrac{\sum_{i=1}^r p_i}{\sum_{i=1}^r q_i} \geq \dfrac{p_r}{q_r}
\end{equation}
so we consider $k=r+1$
\begin{align}
    \begin{split}
        \dfrac{\sum_{i=1}^r p_i}{\sum_{i=1}^r q_i} & \geq \dfrac{p_{r+1}}{q_{r+1}},                                                       \\
        \implies \dfrac{\sum_{i=1}^r p_i}{\sum_{i=1}^r q_i} & \geq \dfrac{\sum_{i=1}^{r+1} p_i}{\sum_{i=1}^{r+1} q_i}\geq\dfrac{p_{r+1}}{q_{r+1}}. \\
    \end{split}
\end{align}
Therefore, we can write the following
\begin{equation}
    \dfrac{p_1+p_2}{q_1+q_2}\geq \dfrac{\sum_{i=1}^r p_i}{\sum_{i=1}^r q_i} \geq \dfrac{\sum_{i=1}^{N} p_i}{\sum_{i=1}^{N} q_i},
\end{equation}
which concludes the proof.\qedsymbol

We will prove the main statement in Eq. (\ref{eq:entropy_maj}) by contradiction. We assume the following,
\begin{equation}\label{eq:proof_beta_contra}
    \beta_1 \geq \beta_2
\end{equation}
and try to show $S(\rho_{\beta_1}) \geq S(\rho_{\beta_2})$.
Note, we have consider the energy eigenvalues to be in non-decreasing sequence, \textit{i.e.}, $E_1\leq E_i \leq E_N$. Hence, we write the following (from Eq. (\ref{eq:proof_beta_contra})):
\begin{align}\label{eq:proof_zbeta}
    e^{\beta_1 E_i} \geq                  & e^{\beta_2 E_i}, ~~~~~ \forall E_i, \notag \\
    \implies  e^{-\beta_1 E_i} \leq       & e^{-\beta_2 E_i}, \notag                   \\
    \implies \sum_i e^{-\beta_1 E_i} \leq & \sum_i e^{-\beta_2 E_i}, \notag            \\
    \implies Z(\beta_1) \leq              & Z(\beta_2).
\end{align}
To proceed further, we need to prove our main theorem as under.
\begin{theorem}\label{th:majorized_rho}
    If $\beta_1 \geq \beta_2$, the density matrix $\rho_{\beta_1}$ majorizes $\rho_{\beta_2}$ ($\rho_{\beta_1} \succ \rho_{\beta_2}$). That is
    \begin{equation}\label{eq:proof_maj_rho_the}
        \sum_{i=1}^k \dfrac{e^{-\beta_1 E_i}}{Z(\beta_1)} \geq \sum_{i=1}^k \dfrac{e^{-\beta_2 E_i}}{Z(\beta_2)},
    \end{equation}
    where $k=1(1)N$ and equality holds for $k=N$.
\end{theorem}
\noindent \textit{Proof:} We first note that, since $E_1\leq E_i\leq E_N$, $\frac{e^{-\beta E_1}}{Z(\beta)}\geq \frac{e^{-\beta E_i}}{Z(\beta)} \geq \frac{e^{-\beta E_N}}{Z(\beta)}$. Let us define the ratio
\begin{equation}\label{eq:proof_th_ratio}
    \mathrm{R}_k = \dfrac{\sum_{i=1}^k e^{-\beta_1 E_i}/Z(\beta_1)}{\sum_{i=1}^k e^{-\beta_2 E_i}/Z(\beta_2)} = \dfrac{\sum_{i=1}^k e^{-\beta_1 E_i}}{\sum_{i=1}^k e^{-\beta_2 E_i}}\cdot\dfrac{Z(\beta_2)}{Z(\beta_1)}.
\end{equation}
Consider the case $k=1$,
\begin{align}
    \begin{split}
        \mathrm{R}_1 & = \dfrac{e^{-\beta_1 E_1}/Z(\beta_1)}{e^{-\beta_2 E_1}/Z(\beta_2)},         \\
       \implies \qquad              & = \dfrac{\sum_j e^{-\beta_2 (E_j - E_1)}}{\sum_i e^{-\beta_1 (E_i - E_1)}}.
    \end{split}
\end{align}
Given, $E_1\leq E_i, ~~ \forall i$, we can write using Eq (\ref{eq:proof_beta_contra}),
\begin{align}
    e^{-\beta_2(E_i-E_1)}                & \geq e^{-\beta_1(E_i-E_1)},\notag                               \\
  \implies  \sum_j e^{-\beta_2(E_j-E_1)}         & \geq \sum_i e^{-\beta_1(E_i-E_1)},\notag                        \\
   \implies \mathrm{R}_1                         & \geq 1,\label{eq:proof_th_r1}                                   \\
   \implies  \dfrac{e^{-\beta_1 E_1}}{Z(\beta_1)} & \geq \dfrac{e^{-\beta_2 E_1}}{Z(\beta_2)} \label{eq:proof_th_1}
\end{align}
Let us consider the ratio $\mathrm{R}_2$
\begin{equation}
    \mathrm{R}_2 = \dfrac{\left(e^{-\beta_1 E_1}+e^{-\beta_1 E_2}\right)/Z(\beta_1)}{\left(e^{-\beta_2 E_1}+e^{-\beta_2 E_2}\right)/Z(\beta_2)}.
\end{equation}
Using Lemma \ref{lemma:ratio} we have:
\begin{align}
    \begin{split}
        \dfrac{\left(e^{-\beta_1 E_1}+e^{-\beta_1 E_2}\right)}{\left(e^{-\beta_2 E_1}+e^{-\beta_2 E_2}\right)}                                    & \geq \dfrac{Z(\beta_1)}{Z(\beta_2)}, \\
        \implies \dfrac{\left(e^{-\beta_1 E_1}+e^{-\beta_1 E_2}\right)}{\left(e^{-\beta_2 E_1}+e^{-\beta_2 E_2}\right)}\cdot\dfrac{Z(\beta_2)}{Z(\beta_1)} & \geq 1,                              \\
    \end{split}
\end{align}
thus satisfying the second step of this proof by induction
\begin{equation}
    \mathrm{R}_2 \geq 1.
\end{equation}
Similarly, we consider the next step, and assume $\mathrm{R}_r\geq 1$ for some $1\leq r\leq N-1$, which implies
\begin{equation}
    \dfrac{\sum_{i=1}^r e^{-\beta_1 E_i}}{\sum_{i=1}^r e^{-\beta_2 E_i}}\cdot\dfrac{Z(\beta_2)}{Z(\beta_1)}\geq 1
\end{equation}
Using Lemma \ref{lemma:ratio} once more, we can write
\begin{align}
    \begin{split}
        \dfrac{\sum_{i=1}^r e^{-\beta_1 E_i}}{\sum_{i=1}^r e^{-\beta_2 E_i}}                                    & \geq\dfrac{\sum_{i=1}^r e^{-\beta_1 E_i}}{\sum_{i=1}^r e^{-\beta_2 E_i}} \geq \dfrac{Z(\beta_1)}{Z(\beta_2)}, \\
      \implies  \dfrac{\sum_{i=1}^r e^{-\beta_1 E_i}}{\sum_{i=1}^r e^{-\beta_2 E_i}}                                    & \geq \dfrac{Z(\beta_1)}{Z(\beta_2)},                                                                          \\
        \implies \dfrac{\sum_{i=1}^r e^{-\beta_1 E_i}}{\sum_{i=1}^r e^{-\beta_2 E_i}}\cdot\dfrac{Z(\beta_2)}{Z(\beta_1)} & \geq 1,                                                                                                       \\
        \implies \mathrm{R}_{r+1}\geq 1
    \end{split}
\end{align}
So, we conclude that the ratio in Eq. (\ref{eq:proof_th_ratio}) is always greater than or equal to one for the desired values of $k=1(1)N$ while being strictly equal to one at $k=N$ implying $\rho_{\beta_1} \succ \rho_{\beta_2}$. This proves the theorem. \qedsymbol \\

Now we can use Schur concavity property \cite{marshall_2011_inequalities} to claim since $\rho_{\beta_1} \succ \rho_{\beta_2}$, $S(\rho_{\beta_1}) \leq S(\rho_{\beta_2})$. But this contradicts the claim in Eq. (\ref{eq:entropy_maj}). Thus our original assumption, $\beta_1 \geq \beta_2$ was wrong. And we have $S(\rho_{\beta_1}) \geq S(\rho_{\beta_2}) \implies \beta_{1} \leq \beta_{2}$. In the same spirit, we note that, given the result of Theorem \ref{th:majorized_rho}, \(\beta_1 \geq \beta_2 \implies \rho_{\beta_1} \succ \rho_{\beta_2}\), and we have the same relation with \(\rho_\beta\)
and \(S(\rho_\beta)\) using Schur's concavity property, we see the reverse is also true. That is, \(\beta_1 \leq \beta_2 \implies S(\rho_{\beta_1}) \geq S(\rho_{\beta_2})\). Therefore, we prove our original claim.

I thank Dr. Budhaditya Bhattacharjee for helpful discussions, and I acknowledge the financial support from the Institute for Basic Science (IBS) in the Republic of Korea through Project No. IBS-R024-D1.

\bibliography{reference}

\end{document}